\documentclass[10pt,preprint]{aastex}

\newcommand{\avg}[1]{{\langle{#1}\rangle}}

\def\simless{\mathbin{\lower 3pt\hbox
	{$\,\rlap{\raise 5pt\hbox{$\char'074$}}\mathchar"7218\,$}}} 
\def\simgreat{\mathbin{\lower 3pt\hbox
	{$\,\rlap{\raise 5pt\hbox{$\char'076$}}\mathchar"7218\,$}}} 

\newcounter{thefigs}
\newcommand{\fignum}{\arabic{thefigs}}

\newcounter{thetabs}

\newcommand{\EW}{W}

\begin{document}
 

\title{The Distribution of [OII] Emission-line Widths of LCRS Galaxies}


\author{Michael Blanton}
\affil{NASA/Fermilab Astrophysics Center\\
Fermi National Accelerator Laboratory, Batavia, IL 60510-0500;\\
blanton@fnal.gov}
\author{\and Huan Lin\footnote{Hubble Fellow}}
\affil{Steward Observatory, University of Arizona,\\
933 N.~Cherry Ave., 
Tucson, AZ 85721; \\
hlin@as.arizona.edu}


\begin{abstract}
We present a simple functional form for the joint distribution of
$R$-band luminosity and [OII] 3727 emission-line equivalent widths of
galaxies, and show that this form is a good fit to the galaxies in the
Las Campanas Redshift Survey. We find a relationship between [OII]
equivalent width $W$ and $R$-band luminosity $L_R$ of the approximate
form: $\avg{W}\approx (10\AA) (L_R/L_{R,\ast})^{-1/2}$, where
$L_{R,\ast}$ is the characteristic luminosity in the Schechter
function. Because this joint distribution yields information about the
relationship between stellar mass in a galaxy and its recent
star-formation rate, it can be useful for testing theories of galaxy
formation. Furthermore, understanding this joint distribution locally
will make it easier to interpret the evolution of [OII] emission-line
widths to higher redshifts.
\end{abstract}


%
%

\section{Motivation}
\label{motiv}

Modern redshift surveys such as the Las Campanas Redshift Survey
(LCRS; \citealt{shectman96a}) have large, homogeneous sets of spectra
from which one can measure star-formation indicators such as the [OII]
3727 \AA\  forbidden line.  It is known that the luminosity function of
galaxies is dependent on the emission-line properties of the galaxies
under consideration (\citealt{lin96a}; \citealt{cowie96a};
	\citealt{ellis96a}; \citealt{small97a}) but the detailed relationship 
between these emission-line properties and galaxy luminosity has not
been explored.  Here we present a calculation of the joint
distribution of $R$-band luminosity and the equivalent width of the
[OII] 3727 line for LCRS galaxies, as well as an analytic form for
this distribution which fits the data well. This joint distribution is
a useful quantity to compare with the predictions of galaxy formation
models ({\it e.g.}, \citealt{cen98a},
\citealt{pearce99a},
\citealt{somerville99a}, \citealt{kauffmann99a}).

While the LCRS is the largest completed redshift survey to date, there
are at least three drawbacks to the sample to be kept in mind.  First,
the survey is $R$-band selected and limited by central surface
brightness; thus, the latest type galaxies, which typically have the
strongest emission lines, are preferentially excluded from the survey,
potentially biasing our results. Second, the fits to the equivalent
widths of the emission lines fail for some galaxies, because their
spectra do not have sufficient signal-to-noise to measure the line. It
is likely that the failure rate of the fit depends on the true
equivalent width of the line, and this unknown incompleteness is a
potential worry. On the other hand, we show below that our results are
robust to the lower limit of equivalent widths we consider.  Given the
typical equivalent width errors of $2\AA$, our results are most
appropriate for galaxies with equivalent widths $> 4\AA$, to which we
limit our sample. Redshift surveys underway, such as the Sloan Digital
Sky Survey (SDSS;
\citealt{york00a}) and the Two-degree Field Galaxy Redshift Survey
(2dFGRS;
\citealt{colless98a}), will be able to overcome these two
difficulties. A final problem, noted by \citet{kochanek00a}, is that
the spectra are taken using fibers with a diameter of $\sim 3''$,
which for typical distances of galaxies in the sample is about 4
$h^{-1}$ kpc. This may cause an ``aperture bias'' which underestimates
the equivalent width of emission lines at low redshift because the
fiber probes the inner, bulge component of spirals, rather than their
disks, which contain the bulk of the star-formation. The SDSS may be
able to constrain this effect by examining the four optical colors
which the survey will measure, and comparing the colors within
fiber-sized apertures to the global colors of each galaxy.

This paper is organized as follows. Section \ref{method} briefly
describes our method for calculating the joint luminosity and
equivalent width function, and presents a simple fitting function
based on that of
\citet{schechter76a}. Section
\ref{results} describes the results using LCRS $R$-band luminosities
and equivalent widths of [OII] 3727 measured by \citet{lin96a}.
Section \ref{conclusions} suggests directions of future research.

\section{Joint Distribution of Luminosity and Equivalent Width}
\label{method}

We follow \citet{sandage79a} and \citet{efstathiou88a},
maximizing the conditional probability that each galaxy $j$, given its
redshift $z_j$,  has its
measured luminosity $L_j$ and equivalent width $W_j$:
\begin{eqnarray}
\label{conditional}
p(L_{j}, \EW_j | z_j) &=& \frac{p(L_j, \EW_j ,z_j)}{p(z_j)} \cr
&=& \frac{\Phi(L_j,
\EW_j)f_g(m_j)}
{\int_{L_{{\mathrm{min}}(z_j)}}^{L_{{\mathrm{max}}(z_j)}} dL
\int_{\EW_{{\mathrm{min}}}}^{\EW_{{\mathrm{max}}}} d\EW
\Phi(L,\EW) f_g(m)},
\end{eqnarray}
Here $L_{{\mathrm{min}}(z_j)}$ and $L_{{\mathrm{max}}(z_j)}$ are the minimum
and maximum luminosities observable at redshift $z_j$, given the flux
limits of the field which contains galaxy
$j$. $\EW_{{\mathrm{min}}}$ and $\EW_{{\mathrm{max}}}$ are the
minimum and maximum values of the equivalent widths of our
sample. (\citealt{lin96a} find the minimum observable equivalent width
to be approximately constant with redshift). $f_g(m)$ represents the
magnitude dependence of the redshift completeness.  The likelihood of
a given model for $\Phi(L,\EW)$ is given by the product of this
conditional probability over all galaxies in the sample. Since this
conditional likelihood is independent of density, the normalization
must be calculated separately.  We use the simple estimator:
\begin{equation}
n_1 = \frac{1}{V} \sum_{j=1}^{N_{\mathrm{gals}}} \frac{1}{\phi(z_j)},
\end{equation}
where $V$ is the size of the volume probed, and $\phi(z)$ is the
selection function:
\begin{equation}
\phi(z) = 
\int_{L_{\mathrm{min}}(z)}^{L_{\mathrm{max}}(z)} dL\,
\int_{\EW_{\mathrm{min}}}^{\EW_{\mathrm{max}}}
d\EW\,
 \Phi(L,\EW) f_g(m) f_t .
\end{equation} 
$f_t$ is the local sampling fraction.\footnote{For a fuller
explanation of the meaning of the quantities $f_g$ and $f_t$, consult
\citet{lin96a}.} \citet{lin96a} find for the luminosity function that
this estimator yields similar results to the minimum variance
estimator of \citet{davis82a} for this sample. 

We use two models to describe $\Phi(L,\EW)$. First, we use the
non-parametric form described by \citet{efstathiou88a}, whose
extension to the two dimensional plane of $L$ and $\EW$ is
trivial. Essentially, this method divides the $(L,W)$ plane into bins
of equal logarithmic width, and assumes the distribution within each
bin is constant. A fast iterative method can then find the set of
values which maximize the likelihood, and we can estimate the errors
by evaluating the second derivatives of the likelihood function at the
fitted values. 

Second, following \citet{sandage79a}, we find the maximum likelihood
fit to a parametrized function. To do so, we parametrize the joint
function as a modified Schechter function, which is motivated by the
results below:
\begin{equation}
\label{modschechter_eq}
\Phi(L,\EW) dL d\EW = \phi_\ast
\left(\frac{L}{L_\ast}\right)^{\alpha}
\exp\left(-L/L_\ast\right) \Psi(\EW | L)
d\EW
\frac{dL}{L_\ast}
\end{equation}
where the conditional equivalent width function is:
\begin{equation}
\Psi(\EW | L) d\EW =
\frac{1}{\sqrt{2\pi}\sigma_{\EW}} 
\frac{d\EW}{\EW} 
\exp\left[-\frac{1}{2\sigma_{\EW}^2}
\left(\ln\frac{\EW}{\EW_0} - A
\ln\frac{L}{L_\ast}+\frac{\sigma_{\EW}^2}{2}\right)^2
\right] 
\end{equation}
That is, at each luminosity, the equivalent widths are distributed
log-normally about a mean value which can be expressed as a function
of luminosity as:
\begin{equation}
\avg{\EW} = \EW_0
\left(\frac{L}{L_\ast}\right)^A.
\end{equation}
$\sigma_W$ parametrizes the width of the log-normal distribution.  We
use this function, and maximize the likelihood in Equation
(\ref{conditional}) over the five parameters $L_\ast$, $\alpha$,
$\EW_0$, $\sigma_{\EW}$, and $A$. 

For the parametric fit, we calculate the error bars using 200 Monte
Carlo realizations. For each realization, we take the redshifts of all
the galaxies in the actual LCRS sample to be the redshifts for the
``galaxies'' in our realization. Then, we select a luminosity and
[OII] equivalent width for each galaxy using Equation
(\ref{modschechter_eq}), limiting the range of absolute luminosities
for each galaxy to that which is within the flux limits at that
redshift. Then we maximize the likelihood for this realization. This
procedure allows us to examine the distribution of the parameters over
all the realizations, and thus calculate the error bars, and to
determine whether our method is biased. We are also able to directly
compare the likelihood values for the realization to the likelihood
value of the data sample. If the fit is consistent with the data,
these likelihoods should be comparable; if the fit is not consistent,
the likelihood value for the data will always be smaller than that for
the realizations.

We calculate distance moduli assuming an Einstein-de Sitter
universe. We use $K$-corrections of the form $K(z)=2.5
\log_{10}(1+z)$ (\citealt{lin96a}). 
Throughout, we assume $H_0 = 100$ $h$ km/s/Mpc with $h=1$; to convert
to other values of $h$, the absolute magnitude scale is shifted by $5
\log_{10} h$, and the luminosity function normalization by $h^3$. For
plotting purposes we show the luminosity function expressed per unit
logarithm $\hat\Phi(L,\EW) = n_1 (\ln 10) L \Phi(L,\EW)$.

\section{Results for the LCRS}
\label{results}

Here we present the joint distribution of $L_R$ and the equivalent
width of $[\mathrm{OII}]$ 3727, for a sample of galaxies with
$-22.5<M_R<-16.5$ and $5,000$ km/s $< cz< 50,000$ km/s, selected from
the North and South 112-fiber fields in the LCRS. The equivalent
widths were measured by \citet{lin96a}, by fitting for the position,
the width, and the amplitude of a Gaussian to the continuum-subtracted
spectrum near the predicted location of [OII] based on the
redshift. For about 25\% of the objects, the spectra were too low
signal-to-noise to constrain these parameters; the equivalent-width
dependence of this incompleteness is unknown. The estimated errors in
the measured equivalent widths are about $2\AA$ on average.  To
minimize the effects of incompleteness and errors, we include only
measured equivalent widths $> 4 \AA$ in our analysis, leaving about
8,500 galaxies in our sample.

\begin{figure}
\figurenum{\fignum}
\plotone{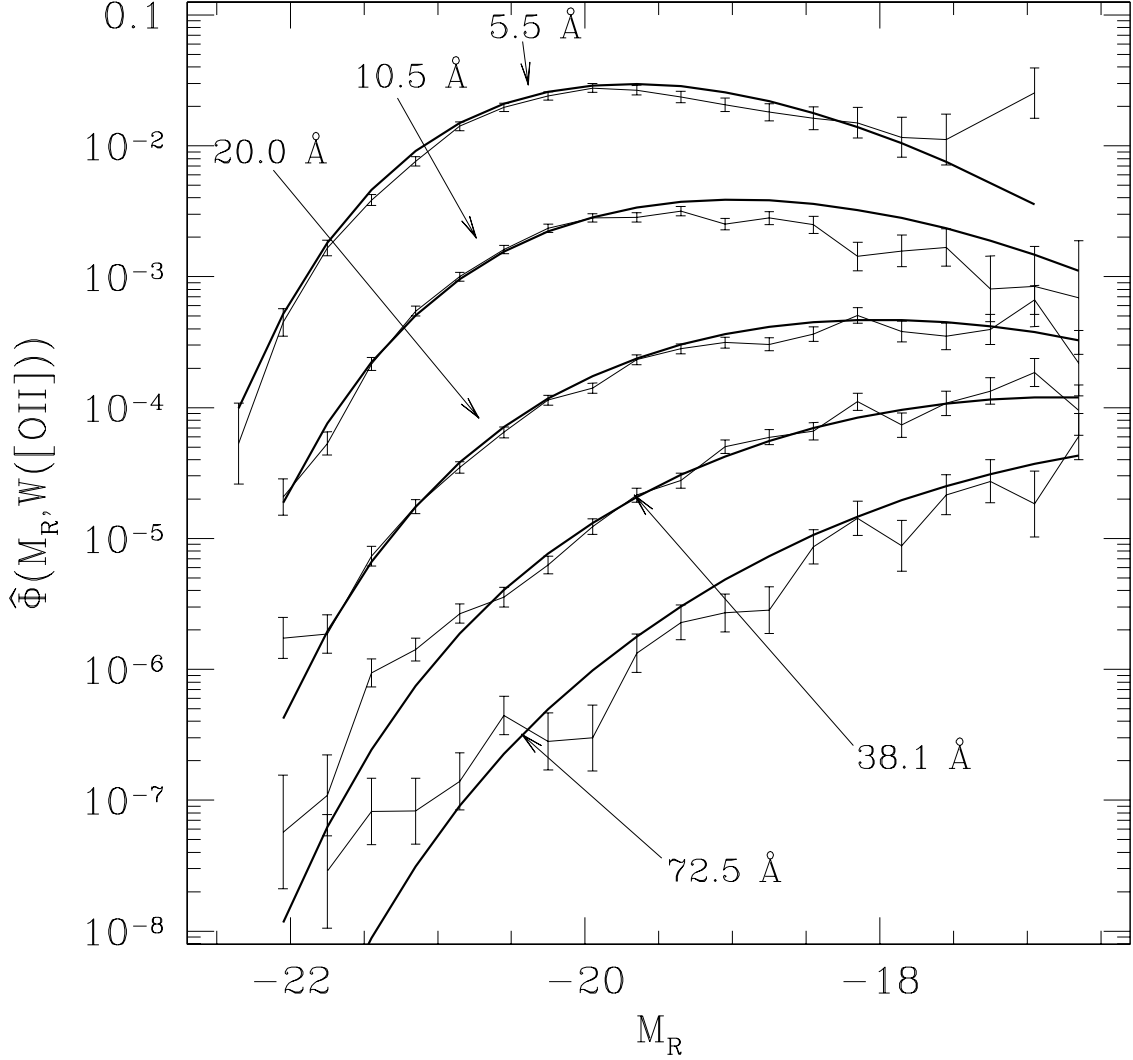}
\caption{\label{lewf_phi} Joint distribution of luminosity and [OII]
equivalent width for the approximately 8,500 LCRS galaxies (in the
N112 and S112 fields) for which we have measured equivalent widths in
the range $4$--$100$ $\AA$. Curves with error bars represent the
results of a two-dimensional non-parametric fit based on the method of
\citet{efstathiou88a}. They are labeled by the central value of each
logarithmically spaced bin in equivalent width. Note the
characteristic differences in $M_\ast$ and faint-end slope between the
star-forming, high equivalent width galaxies, and the quiescent, low
equivalent width galaxies. Smooth curves represent the best fit
modified Schechter function of Equation (\ref{modschechter_eq}), which
appears to model the data well. Parameters of this fit as well as
their error bars and covariances are given in Table
\ref{lewf_phi_table}. For the purposes of clarity, we have offset the
$5.5 \AA$, $10.5 \AA$, and $20.0 \AA$ curves (for both the
non-parametric and the modified Schechter fits) by 1.8, 1.0, and 0.3
dex, respectively. }
\end{figure}

Figure \ref{lewf_phi} shows the non-parametric fit as the thin solid
lines with error bars. Each line shown represents a bin of equivalent
widths, whose central value is given. Some of the lines are offset for
clarity, as described in the caption. Note the characteristic
difference between the strong emission line galaxies, which are in
general less luminous and have a steeper faint-end slope, and the weak
emission line galaxies, which are brighter with a shallower faint-end
slope. This result accords qualitatively with that of \citet{lin96a}
and those of numerous other investigations of the dependence of the
luminosity function on [OII] equivalent width (\citealt{cowie96a};
\citealt{ellis96a}; \citealt{small97a}) and 
on spectral type in general (\citealt{zucca97a};
\citealt{bromley98a}; \citealt{folkes99a}; \citealt{loveday99a}).

We also show the modified Schechter function fit in Figure
\ref{lewf_phi} as the thick solid lines for each equivalent width
shown (again, some are offset for clarity).  Apparently this model
does a pretty good job, though it uses 6 parameters: the ordinary
Schechter parameters $\phi_\ast$, $L_\ast$, $\alpha$, plus the
parameters describing the dependence of equivalent width on luminosity
$\EW_0$, $\sigma_{\EW}$, and $A$ (which is negative, because brighter
galaxies have smaller equivalent widths). The best-fit values of these
parameters are given in Table
\ref{lewf_phi_table}. Note that it is approximately true from these 
results that
\begin{equation}
\avg{\EW} \approx (10 \AA) \left(\frac{L_R}{L_{R,\ast}}\right)^{-1/2}.
\end{equation}
Also, the Schechter parameters $\phi_\ast$, $L_\ast$ and $\alpha$
agree generally with the results of the independent analysis of
\citet{lin96a}, although the faint end slope here is a bit
steeper. 

Table \ref{lewf_phi_table} also gives the errors in the modified
Schechter parameters, as well as the correlation matrix between these
parameters, determined from 200 Monte Carlo realizations, as described
above. We have found that the bias in the maximum likelihood method is
smaller than the error bars in this sample. Furthermore, we find that
the fraction of Monte Carlo realizations which have likelihoods worse
than that found for the data is about $P_{\mathrm{worse}}\approx
0.47$; this means that the likelihood for the data is comparable to
the likelihoods from the Monte Carlo realizations, indicating that the
fit is consistent with the data. 


We have experimented with performing the modified Schechter function
fit with limiting equivalent widths between $0.5\AA$ and $9.5\AA$,
instead of the limiting value of $4 \AA$ used for the results
just presented. The parameters appear fairly robust to what this lower
limit is.  The largest changes are in the faint-end slope, which
varies from $\alpha\sim -0.75$ at a limiting equivalent width of $0.5
\AA$ to $\alpha\sim -1.1$ at a limiting equivalent width of $9.5
\AA$; that we measure a slightly different faint-end slope than
\citet{lin96a} is thus related to our choice of a limiting equivalent
width of $4\AA$. Meanwhile, $M_\ast$ varies by about 0.15
magnitudes. However, the changes in the parameters which describe the
distribution of equivalent widths are quite small. $\EW_0$ varies by
$<3\%$, $\sigma_\EW$ varies from $0.85$ to $0.75$, and $A$ varies from
$-0.45$ to $-0.49$. This consistency simply tells us that the modified
Schechter function is good at fitting the combination of the intrinsic
equivalent width distribution and the incompleteness as a function of
equivalent width.  Nevertheless, we find it encouraging that the
nearly same analytic form fits equally well the high equivalent width
galaxies, which we are fairly confident of, and the low equivalent
width galaxies, which may suffer from incompleteness as a function of
equivalent width.

\section{Discussion}
\label{conclusions}

We have presented a simple functional form which seems to describe
well the joint distribution of luminosity and the equivalent width of
the [OII] 3727 emission line in the LCRS. We caution that the
dependence of completeness on equivalent width is unknown, and
further, that we have not accounted for the distribution of the
equivalent width errors (on average about $2 \AA$) in our analysis.
Upcoming surveys such as the SDSS and 2dFGRS will provide larger
homogeneous sets of spectra with better resolution, and will overcome
a number of the problems encountered here.  

The joint distribution function $\Phi(L_R,W)$ can provide a useful
tool for testing theories of galaxy formation, because the $R$-band
luminosity is an approximate indication of the stellar mass contained
in each galaxy and the equivalent width of [OII] 3727 is an
approximate indication of recent star-formation in the galaxy. By
combining hydrodynamic or semi-analytic models for galaxy formation,
such as those mentioned above, with spectral synthesis models
(\citealt{leitherer96a}; \citealt{kennicutt98a}), it may be possible
to place strong constraints on the properties of the star-formation
history of galaxies. In this vein, understanding this joint
distribution locally is also helpful in interpreting the evolution of
[OII] emission at higher redshifts and thus the evolution of the
star-formation rate of the universe (\citealt{hogg98a}).

\acknowledgments

Thanks to Scott Dodelson, Daniel Eisenstein, David Hogg, Siang Peng
Oh, Ravi Sheth, Doug Tucker, and Idit Zehavi for useful discussions.
Thanks to Michael Strauss for extensive advice and comments. MB is
grateful for the hospitality of the Department of Physics and
Astronomy at the State University of New York at Stony Brook, who
kindly provided computing facilities on his frequent visits there.  MB
acknowledges the support of the DOE and NASA grant NAG 5-7092 at
Fermilab.  HL acknowledges support provided by NASA through Hubble
Fellowship grant \#HF-01110.01-98A awarded by the Space Telescope
Science Institute, which is operated by the Association of
Universities for Research in Astronomy, Inc., for NASA under contract
NAS 5-26555. Finally, this work would not have been possible without
the public availability of the Las Campanas Redshift Survey data, for
which we thank the LCRS team.



\newpage

\begin{deluxetable}{crrrrr}
\tablewidth{0pt}
\tablecolumns{6}
\tablecaption{\label{lewf_phi_table} Modified Schechter Fit to LCRS
Galaxies}
\tablecomments{Parameters of the modified Schechter function given in
Equation (\ref{modschechter_eq}) and correlation matrix between the
parameters.  The first line of the table gives the values of the
parameters and their errors. The bottom section of the table gives the
correlation matrix. Normalization is $\phi_\ast = 1.34 \pm 0.03$
($\times 10^{-2}$) Mpc $^{-3}$. The errors and the correlation matrix
were calculated using 200 Monte Carlo simulations. The fraction of
realizations which had worse fits to the model than did the data was
about $P_{\mathrm{worse}}=0.47$, indicating that the model is
consistent with the data.}
\tablehead{ & $M_\ast$ & $\alpha$ & \EW$_0$ ($\AA$) &
$\sigma_{\mathrm{\EW}}$ & A }
\startdata
 & $ -20.32 \pm  0.01$  & $ -0.91 \pm  0.02$  & $ 10.14 \pm  0.03$  & $  0.77 \pm  0.01$  & $ -0.47 \pm  0.01$ \cr
\hline
\hline\cr
$M_\ast$ &  1.00 &  0.39 & -0.69 &  0.16 & -0.25 \cr
$\alpha$ &  0.39 &  1.00 & -0.73 &  0.28 & -0.32 \cr
$W_0$ & -0.69 & -0.73 &  1.00 & -0.34 &  0.11 \cr
$\sigma_W$ &  0.16 &  0.28 & -0.34 &  1.00 & -0.45 \cr
$A$ & -0.25 & -0.32 &  0.11 & -0.45 &  1.00 \cr
\enddata
\end{deluxetable}

\clearpage

\setcounter{thefigs}{0}

\newpage
\stepcounter{thefigs}

\end{document}